\documentclass[12pt,epsf,amstex]{article}
\usepackage [dvips]{graphicx}
\usepackage{amsmath}
\usepackage{amssymb}
\usepackage{epsfig}

\addtocounter{secnumdepth}{1}
\begin{document}

\newcommand{\bP}{\bar{P}}
\newcommand{\br}{\bar{r}}
\newcommand{\bv}{\bar{v}}
\newcommand{\bz}{\bar{z}}
\newcommand{\Kc}{K_{\rm c}}
\newcommand{\Tc}{T_{\rm c}}
\newcommand{\ms}{m_{\rm sp}}
\newcommand{\bbeta}{\bar{\beta}}
\newcommand{\bgamma}{\bar{\gamma}}
\newcommand{\btau}{\bar{\tau}}
\newcommand{\bomega}{\bar{\omega}}
\newcommand{\bOmega}{\bar{\Omega}}
\newcommand{\tbeta}{\tilde{\beta}}
\newcommand{\tgamma}{\tilde{\gamma}}
\newcommand{\trho}{\tilde{\rho}}
\newcommand{\ttau}{\tilde{\tau}}
\newcommand{\tOmega}{\tilde{\Omega}}
\newcommand{\td}{\tilde{d}}
\newcommand{\tN}{\tilde{N}}
\newcommand{\tT}{\tilde{T}}
\newcommand{\rhotot}{{\rho}_{\rm tot}}
\newcommand{\rhototlead}{{\rho}_{\rm tot,lead}}
\newcommand{\rhototfoll}{{\rho}_{\rm tot,foll}}
\newcommand{\rholead}{{\rho}_{\rm lead}}
\newcommand{\rhofoll}{{\rho}_{\rm foll}}
\newcommand{\sigmafoll}{{\sigma}_{\rm foll}}
\newcommand{\flead}{f_{\rm lead}}
\newcommand{\ffoll}{f_{\rm foll}}
\newcommand{\brho}{\bar{\rho}}
\newcommand{\brholead}{\bar{\rho}_{\rm lead}}
\newcommand{\brhofoll}{\bar{\rho}_{\rm foll}}
\newcommand{\tomega}{\tilde{\omega}}
\newcommand{\hrho}{\hat{\rho}}
\newcommand{\htau}{\hat{\tau}}

\newcommand{\bE}{{\bf{E}}}
\newcommand{\bO}{{\bf{O}}}
\newcommand{\bR}{{\bf{R}}}
\newcommand{\bS}{{\bf{S}}}
\newcommand{\bT}{\mbox{\bf T}}
\newcommand{\bt}{\mbox{\bf t}}
\newcommand{\half}{\frac{1}{2}}
\newcommand{\thalf}{\tfrac{1}{2}}
\newcommand{\bsA}{\mathbf{A}}
\newcommand{\bsV}{\mathbf{V}}
\newcommand{\bsE}{\mathbf{E}}
\newcommand{\bsT}{\mathbf{T}}
\newcommand{\bsZ}{\hat{\mathbf{Z}}}
\newcommand{\bse}{\mbox{\bf{1}}}
\newcommand{\bspsi}{\hat{\boldsymbol{\psi}}}
\newcommand{\cdottt}{\!\cdot\!}
\newcommand{\deltaR}{\delta\mspace{-1.5mu}R}
\newcommand{\invup}{\rule{0ex}{2ex}}

\newcommand{\bGamma}{\boldmath$\Gamma$\unboldmath}
\newcommand{\dd}{\mbox{d}}
\newcommand{\ee}{\mbox{e}}
\newcommand{\p}{\partial}

\newcommand{\rmax}{r_{\rm max}}

\newcommand{\artanh}{\mbox{artanh}}
\newcommand{\wrj}{w^{r}_j}
\newcommand{\wrzerj}{w^{r}_{0,j}}
\newcommand{\wronej}{w^{r}_{1,j}}
\newcommand{\wrtwoj}{w^{r}_{2,j}}
\newcommand{\wsj}{w^{s}_j}
\newcommand{\Wrj}{W^{r}_j}
\newcommand{\Wsj}{W^{s}_j}
\newcommand{\Wsi}{W^{s}_i}
\newcommand{\wGj}{w^G_j}
\newcommand{\wstarrj}{w^r_{*,j}}
\newcommand{\wstarsj}{w^s_{*,j}}
\newcommand{\Pst}{P^{\rm st}}
\newcommand{\la}{\langle}
\newcommand{\ra}{\rangle}
\newcommand{\rao}{\rangle\raisebox{-.5ex}{$\!{}_0$}}  
\newcommand{\rae}{\rangle\raisebox{-.5ex}{$\!{}_1$}}
\newcommand{\beq}{\begin{equation}}
\newcommand{\eeq}{\end{equation}}
\newcommand{\bea}{\begin{eqnarray}}
\newcommand{\eea}{\end{eqnarray}}
\def\lsim{\:\raisebox{-0.5ex}{$\stackrel{\textstyle<}{\sim}$}\:}
\def\gsim{\:\raisebox{-0.5ex}{$\stackrel{\textstyle>}{\sim}$}\:}

\numberwithin{equation}{section}

\thispagestyle{empty}
\title{\Large {\bf 
Note on a $q$-modified central limit theorem\\ [5mm]  
\phantom{xxx} }} 
 
\author{{H.J.~Hilhorst}\\[5mm]
{\small Laboratoire de Physique Th\'eorique, B\^atiment 210}\\[-1mm] 
{\small Universit\'e Paris-Sud 11 and CNRS,
91405 Orsay Cedex, France}\\[5mm]}

\maketitle
\begin{small}
\begin{abstract}
\noindent A $q$-modified version of the central limit theorem
due to Umarov {\it et al.} affirms that 
$q$-Gaussians are attractors under addition and rescaling
of certain classes of strongly correlated random variables.
The proof of this theorem rests on a nonlinear
$q$-modified Fourier transform. By exhibiting an invariance property
we show that this Fourier transform does not have an inverse. 
As a consequence, the theorem
falls short of achieving its stated goal.
\vspace{8mm}

\end{abstract}
\end{small}
\vspace{66mm}

\noindent LPT Orsay 10/61
\newpage


\section{Introduction} 
\label{secintroduction}

Since over two decades 
work by Tsallis and co-workers 
\cite{Tsallis88,GellMannTsallis04,Tsallis09} 
has drawn attention to problems in
statistical physics where Boltzmann-Gibbs
statistics is not or not directly applicable.
These problems typically involve long-range interactions and correlations.
Tsallis initiated an approach to such systems that is often referred to as
{\it $q$-modified\,} or  {\it nonextensive\,} statistical mechanics.
One of its characteristics is the occurrence of so-called
$q$-Gaussian probability distributions; these are defined by
\beq
G_{q}(x) 
= \frac{ C_q }
{ \left[1+({q}-1)x^2\right]^{\frac{1}{{q}-1}} }\,,
\label{qGaussian}
\eeq
where $q$ is an arbitrary real number, $C_q$ is the normalization constant, 
and the domain of
definition is the real $x$-axis (for $q\geq 1$) or  
the part thereof where $1+(q-1)x^2 \geq 0$ (for $q<1$). 
For $q \to 1$ equation (\ref{qGaussian}) reduces to an ordinary Gaussian. 
\vspace{3mm}

In $q$-modified statistical mechanics
it is expected that $q$-Gaussians arise naturally
when one considers sums of {\it strongly\,}
correlated variables, at least
for a certain class of correlations; 
this by analogy to ordinary Gaussians, that
describe sums of sufficiently {\it weakly\,} correlated variables.
Therefore, the observation of $q$-Gaussians, whether in nature
or in numerical simulations, would lend support to the
applicability of $q$-modified statistical mechanics.
\vspace{3mm}

There have recently been 
attempts to provide a theoretical basis for
this special status attributed to $q$-Gaussians.
One attempt
has consisted in numerically
generating and adding up random variables
with strong correlations of some well-controlled kind, and determining the
distribution of the sum. 
In two examples studied by Thistleton {\it et al.}
\cite{Tsallis07,Thistletonetal09} and by Moyano {\it et al.} 
\cite{Moyanoetal06} 
the sum distribution is numerically virtually
indistinguishable from a $q$-Gaussian and each was initially believed by
their authors to be one.
In subsequent analytical work Hilhorst and Schehr
\cite{HilhorstSchehr07} showed, however, 
that in fact the analytic expressions of
these sum distributions are unrelated to $q$-Gaussians.
It was pointed out
\cite{HilhorstSchehr07,Hilhorst09}, moreover,
that correlations between 
variables may be engineered in such a way that the
sum of $N$ of them,
scaled and in the limit $N\to\infty$, has any desired distribution.

The question therefore is not whether a sum of strongly
correlated variables {\it can\,} 
have a $q$-Gaussian distribution%
\footnote{This is indeed possible; among the examples 
are those of references \cite{Rodriguezetal08,Hilhorst09}.},
but whether under the operation of addition and scaling of random variables
such $q$-Gaussians appear as {\it attractors,} 
at least for certain classes of correlations
wide enough to be of interest for physics.
\vspace{3mm}

A second attempt to mathematically
consolidate the special role at\-tri\-buted to the
$q$-Gaussian probability law
is due to Umarov {\it et al.} \cite{Umarovetal08}. These authors,
in the wake of earlier conjectures \cite{Tsallis05,Moyanoetal06},
prove a $q$-modified central limit theorem ($q$-CLT) 
in which the limit functions are $q$-Gaussians%
\footnote{Extensions of the theorem 
to the multivariate case  \cite{UmarovTsallis07} and to
$q$-modified $\alpha$-stable L\'evy distributions \cite{Umarovetal10}
have appeared since.}.
We recently expressed concern \cite{Hilhorst09} 
about two points in the proof of this theorem:

(i) First, let $x_1,x_2,\ldots,x_n,\ldots$ denote the sequence of
random variables whose scaled partial sums we wish to study.
Then in order for the $q$-CLT 
to apply, the first $N$ of these variables, for all $N=1,2,\ldots$,
must be correlated according to a certain
condition stated explicitly in \cite{Umarovetal08}
and termed ``$q$-independence''.
This property indeed holds,
according to \cite{Umarovetal08},
for a conveniently chosen sequence of random variables that are themselves,
individually, $q$-Gaussian distributed. 
However, to our knowledge,
no other example 
that would illustrate the theorem
has been exhibited so far.
It is therefore not sure that it is at all possible 
to fulfill the conditions of the theorem in any nontrivial way.

(ii) Secondly, the proof of the theorem is based on the use of a
$q$-modi\-fi\-ca\-tion of the ordinary Fourier transform. This $q$-FT is
nonlinear and, again, tailored to $q$-Gaussian distributions:
when applied to a $q$-Gaussian, it produces (up to scaling factors) a
$q^\prime$-Gaussian where $q^\prime(q)$  
is a known function.
It was observed in reference \cite{Hilhorst09} that this $q$-FT
does not have an inverse.
This observation is the subject of the present note.

We will first show here the proof of
noninvertibility 
hinted at in reference \cite{Hilhorst09} and which is based on a
counterexample.  
Whereas a single counterexample may be mathematically sufficient to make a
point, we next ask
whether one can perhaps obtain an invertible transformation 
by restricting the domain of action of the $q$-FT to a suitable
subspace of probability distributions. The answer turns out to be
negative. 
By explicit construction we exhibit
families of functions 
all having the same $q$-Fourier transform; 
and we show that the $q$-Gaussians themselves are part of such families. 

Sections \ref{secmodified} and \ref{secnoninvert} deal with the
definition and the non-invertibility of the $q$-Fourier transform,
respectively. In section \ref{secconclusion} we present a brief
discussion and conclusion.


\section{A $q$-modified Fourier transform}
\label{secmodified}

Let $f(x)\geq 0$ 
be an integrable function on the real axis. 
Umarov {\it et al.} \cite{Umarovetal08}
define its nonlinear $q$-Fourier transform $\hat{f}_q(\xi)$ as 
\beq
\hat{f}_q(\xi) = \int_{-\infty}^\infty \!\dd x\,
\frac{f(x)}
{ \big[ 1-(q-1)\,{\rm i}\,\xi xf^{q-1}(x)\big]^{ \frac{1}{q-1} } }\,,
\label{defqFT}
\eeq
where $\xi$ is real and from now on $q>1$; furthermore
here and henceforth $f^{q-1}(x) \equiv [f(x)]^{q-1}$.  
For $q\to 1^+$ expression (\ref{defqFT})
reduces to the standard Fourier transform.
If $f(x)$ is a probability distribution, as will be the case,  
the normalization condition  
\beq
\int_{-\infty}^{\infty} \dd x\, f(x)=1
\label{normfx}
\eeq
has to be imposed. 

The $q$-FT (\ref{defqFT}) has since its introduction
been discussed on various occasions 
by the original authors and others
\cite{UmarovTsallis08,UmarovDuarteQueiros10,Tsallisetal09,JaureguiTsallis10,%
Tsallis09b}.
In what follows we will show that it is not invertible,
even when restricted to the space of probability distributions. 


\section{Noninvertibility of the $q$-FT}
\label{secnoninvert}


\subsection{An example}
\label{secexample}

We begin by examining a particular case (\cite{Hilhorst09}, footnote [30]).\\

\noindent
{\bf Example 1.\,\,}
Let us consider 
\beq
f(x)=\left\{
\begin{array}{ll}
\Big( \dfrac{\lambda}{x} \Big)^{ \frac{1}{q-1} } \phantom{XX}
& x\in(a,b),\\[4mm]
\phantom{X}0 & \mbox{otherwise,}
\end{array}
\right.
\label{exm1}
\eeq
where $0<a<b$ and $\lambda>0$. Due to the normalization (\ref{normfx}) 
the constant $\lambda$ can be expressed in terms of the interval end points 
$a$ and $b$, which yields
\beq
\lambda=\Big[ \,\frac{q-1}{q-2}\,
\big( \,b^{\,\frac{q-2}{q-1}}-a^{\,\frac{q-2}{q-1}}\, \big)\, \Big]^{-(q-1)}.
\label{exlambdaab}
\eeq
Substituting (\ref{exm1}) in (\ref{defqFT}) and doing the integral
using the normalization of $f$ leads to
\beq
\hat{f}_q(\xi) = 
\frac{1}{ \big[ 1-(q-1)\,{\rm i}\,\xi\lambda \big]^{ \frac{1}{q-1} } }\,.
\label{resqFT}
\eeq
This is the same result for any interval $(a,b)$ 
that satisfies (\ref{exlambdaab}) with a fixed $\lambda$.
Hence equations (\ref{exm1})-(\ref{exlambdaab}) define a
one-parameter family of normalized functions 
$f(x)$ all having the same $q$-Fourier transform 
(\ref{resqFT}). 
This counterexample 
shows that the $q$-FT is not invertible on the space
of probability distributions.


\subsection{An invariance property}
\label{secgeneralization}

For physicists 
it is important to know that the above example is not an isolated
case that might somehow be eliminated by suitably restricting the space of 
functions.
We will therefore show how
other examples may be constructed,
many of them corresponding to physically reasonable probability 
distributions.
The construction rests on the fact
that for a large class of functions $f(x)$ the expression 
\beq
\lambda(x)=xf^{q-1}(x)
\label{deflambdax}
\eeq
is not invertible to a single-valued function $x(\lambda)$.
Let us rewrite the $q$-FT of equation (\ref{defqFT}) as
\beq
\hat{f}_q(\xi) = \int_{-\infty}^\infty\!\dd\lambda\,
\frac{F(\lambda)}{ \big[ 1-(q-1)\,{\rm i}\,\xi\lambda \big]^{ \frac{1}{q-1} } }
\label{lambdaFT}
\eeq
in which
\bea
F(\lambda) &=& \int_{-\infty}^{\infty}\!\dd x\,\,
\delta\big(xf^{q-1}(x)-\lambda\big)\,f(x) \nonumber\\[2mm]
&=& \sum_i \left|\frac{\dd}{\dd x}[xf^{q-1}(x)]\right|_{x=x_i}^{-1}\,f(x_i),
\label{exFlambda1}
\eea
where the sum runs through the set $\{x_i\}$ of 
the solutions of $xf^{q-1}(x)=\lambda$.
Obviously, if two distinct functions $f_1(x)$ and $f_2(x)$, 
when substituted in
(\ref{exFlambda1}),
lead to the same $F(\lambda)$, then by (\ref{lambdaFT}) they will
have the same $q$-FT. {\it This invariance leads to the nonuniqueness of the
inverse $q$-FT.}


\subsection{A class of symmetric functions}
\label{secspecial}

An investigation of this invariance in its most general
form would probably begin by classifying the $f(x)$
according to the number of terms that they engender in the sum on $i$
in equation (\ref{exFlambda1}).
It is not needed for our purpose to embark on so broad an enterprise.
We will study here
a more limited but
important class of functions $f(x)$ which contains, in
particular, the $q$-Gaussians.
None of the restrictions on $f(x)$ to be adopted below 
is essential; each can be overcome by a little more work.
\vspace{3mm}

Let us consider for convenience a symmetric function, $f(x)=f(-x)$,
that is finite in the origin, $f(0)<\infty$.
Its values for $x>0$ (for $x<0$)
then determine $F(\lambda)$
for $\lambda>0$ (for $\lambda<0$) and we have $F(\lambda)=F(-\lambda)$.
We may therefore limit our analysis to $x, \lambda>0$.
Let us furthermore restrict our attention to those $f(x)$ 
for which $\lambda(x)=xf^{q-1}(x)$ is monotonic
on $(0,\infty)$ except for passing through a single maximum at $x=x_{\rm m}$;
and let $\lambda_{\rm m}\equiv\lambda(x_{\rm m})$.
We note that
the monotonicity condition is not a very severe one and is satisfied, 
in particular, by the 
$q$-Gaussians (\ref{qGaussian}).
\vspace{3mm}

Let now $f(x)$ be in the class delimited above. 
Full generality not being our purpose, let us suppose that $1<q\leq 2$.  
In that case the integrability of $f(x)$ implies that we must have
\beq
\lambda(0)=\lambda(\infty)=0. 
\label{proplambda}
\eeq
Let us denote by $x_-(\lambda)$ and $x_+(\lambda)$ the inverses of
$\lambda(x)$ on the intervals $[0,x_{\rm m}]$ and $[x_{\rm m},\infty)$,
respectively. 
The pair $x_\pm(\lambda)$ is an alternative
representation of $f(x)$.
Using that $\dd x_+/\dd\lambda<0<\dd x_-/\dd\lambda$
and that $f(x)=(\lambda/x)^{\frac{1}{q-1}}$ we obtain from
(\ref{exFlambda1}) for $F(\lambda)$ the expression
\beq
F(\lambda) = \frac{\dd x_-}{\dd\lambda}
\Big(\frac{\lambda}{x_-(\lambda)}\Big)^{\frac{1}{q-1}}\,
-\,\frac{\dd x_+}{\dd\lambda}
\Big(\frac{\lambda}{x_+(\lambda)}\Big)^{\frac{1}{q-1}},
\label{exFlambda2}
\eeq
which we cast in the final form
\beq
F(\lambda) = \tfrac{q-2}{q-1}\lambda^{\frac{1}{q-1}} \frac{\dd}{\dd\lambda}
\Big[ x_-^{\frac{q-1}{q-2}}\,(\lambda) - x_+^{\frac{q-1}{q-2}}(\lambda) \Big]
\label{exFlambdafin}
\eeq
where $0\leq\lambda\leq\lambda_{\rm m}$.
Starting from (\ref{normfx}) and replacing $x$ by $\lambda$ as the variable
of integration, we find along similar lines that the normalization of a
symmetric function $f(x)$ can be expressed as
\beq
2\int_0^{\lambda_{\rm m}}\!\dd\lambda\,F(\lambda) = 1.
\label{normflambda}
\eeq

Equation (\ref{exFlambdafin}) shows that the invariance 
can now be expressed as follows.
The function $F(\lambda)$
does not change if in (\ref{exFlambdafin}) 
we substitute $x_\pm(\lambda) \mapsto \tilde{x}_\pm(\lambda)$ with
\bea
\tilde{x}_-^{\frac{q-1}{q-2}}(\lambda) &=&
x_-^{\frac{q-1}{q-2}}(\lambda)+G(\lambda)+G_-\,, 
\nonumber\\[2mm]
\tilde{x}_+^{\frac{q-1}{q-2}}(\lambda) &=&
x_+^{\frac{q-1}{q-2}}(\lambda)+G(\lambda)+G_+\,,  
\label{subst}
\eea
where $G_+$ and $G_-$ are constants 
and $G(\lambda)$ is an arbitrary function.
Equation (\ref{lambdaFT}) shows that this substitution does
not change the $q$-Fourier transform $\hat{f}_q(\xi)$ and (\ref{normflambda})
shows that it does not change the normalization of $f(x)$.
The function $G(\lambda)$ should satisfy certain rather mild
constraints coming from the fact that we want the pair
$\tilde{x}_\pm(\lambda)$
to be again the representation of a probability distribution $\tilde{f}(x)$. 
This distribution $\tilde{f}(x)$, if it exists, 
can therefore be constructed in the following two steps:
\begin{enumerate}
\item 
Invert $\tilde{x}_\pm(\lambda)$
to a function $\tilde{\lambda}(\tilde{x})$.
\item 
Solve $\tilde{f}(x)$ from $x\tilde{f}^{q-1}(x)=\tilde{\lambda}(x)$,
which gives
\beq
\tilde{f}(x) = \Big(\frac{\tilde{\lambda}(x)}{x}\Big)^{\frac{1}{q-1}}.
\label{exftilde}
\eeq
\end{enumerate}
The result is a function $\tilde{f}(x)$ different from $f(x)$ but which has
the same $q$-FT. 


\subsection{Three more examples}
\label{secexamples}

We will exhibit below three examples that result from an implementation 
of the procedure of section \ref{secspecial}.
We have not sought to exploit the full freedom 
offered by the occurrence of the arbitrary function $G(\lambda)$ in 
(\ref{subst}), but replaced it with a single parameter.\\ 

\noindent
{\bf Example 2.\,\,} Let us take for $f(x)$ the $q$-Gaussian 
$G_q(x)$ of equation (\ref{qGaussian}).
For this function direct calculation gives
\beq
x_\pm(\lambda)=\frac
{C_q^{q-1}\pm\big[C_q^{2(q-1)}-4(q-1)\lambda^2\big]^{\frac{1}{2}}}
{2\lambda(q-1)}\,, \qquad 0<\lambda\leq\lambda_{\rm m}\,,
\label{exxpmlambda}
\eeq
with
\beq
\lambda_{\rm m}=\tfrac{1}{2}(q-1)^{-\frac{1}{2}}C_q^{q-1}. 
\label{exlambdam}
\eeq
Expression (\ref{exxpmlambda}) has the properties
\beq
x_\pm(0)=\left\{
\begin{array}{l}
\infty      \\[2mm]
0 
\end{array}
\right.
\qquad \mbox{ and } \qquad
x_\pm(\lambda_{\rm m})=x_{\rm m}=(q-1)^{-\frac{1}{2}}.
\label{exxm}
\eeq
Let us choose $G(\lambda)=A$ where $A\geq0$ is a parameter,
and $G_+=G_-=0$ in (\ref{subst}).
Then
\beq
\tilde{x}_\pm(\lambda)=
\Big({x}_\pm^{\frac{q-2}{q-1}}(\lambda)+A\Big)^{\frac{q-1}{q-2}}.
\label{xtildeex}
\eeq
In expression (\ref{xtildeex}) 
the $x_\pm(\lambda)$ are given by (\ref{exxpmlambda})
and therefore the $\tilde{x}_\pm(\lambda)$ can be inverted to a function
$\tilde{\lambda}(\tilde{x})$. 
The result is
\beq
\tilde{\lambda}(\tilde{x}) = \frac{2\lambda_{\rm m}y}{1+y^2}\,, \qquad
y\equiv(q-1)^{\frac{1}{2}}
\Big(\tilde{x}^{\frac{q-2}{q-1}}-A\Big)^{\frac{q-1}{q-2}}.
\label{lambdatildextilde}
\eeq
We will now indicate the $A$ dependence explicitly and write 
$\tilde{f}_A(x)$ for the function $\tilde{f}(x)$ represented by 
(\ref{lambdatildextilde}).
Using (\ref{exftilde}) we find from (\ref{lambdatildextilde}) 
\bea
\tilde{f}_A(x) &=& \Big(\frac{\tilde{\lambda}(x)}{x}\Big)^{\frac{1}{q-1}}
\nonumber\\[2mm]
&=& \frac{}{}
\frac{ C_q\,\big(x^{\frac{q-2}{q-1}}-A\big)^{\frac{1}{q-2}} }
{ x^{\frac{1}{q-1}}\,
  \big[ \,1+(q-1)\big(x^{\frac{q-2}{q-1}}-A\big)^{2\frac{q-1}{q-2}}\, 
  \big]^{\frac{1}{q-1}} }\,,
\label{ftildex}
\eea
valid in the domain $A \leq x^{\frac{q-2}{q-1}} <\infty$, 
whereas $\tilde{f}_A(x)=0$ for $0 \leq x^{\frac{q-2}{q-1}} \leq A$
and $\tilde{f}_A(-x)=\tilde{f}_A(x)$. We have not indicated
the $q$-dependence of this family explicitly.
The $A=0$ member of family (\ref{ftildex}) 
is the original $q$-Gaussian $G_q(x)$.
By construction all $\tilde{f}_A(x)$
have the same $q$-FT, independently of $A$.
Hence we have constructed a one-parameter family containing a $q$-Gaussian
and for which the $q$-FT has no inverse.\\


\noindent
{\bf Example 3.\,\,}
We consider the special case $q=\frac{3}{2}$ of equation (\ref{ftildex}).
The reason is that, whereas different statements concerning the $q$-FT
occurring in the literature
may have different domains of validity on the $q$ axis, 
virtually all of them apply
for $1 \leq q \leq 2$. Hence a value in the middle of this interval
is among the most relevant ones.
For $q=\frac{3}{2}$ equation (\ref{ftildex}) simplifies to
\beq
\tilde{f}_A(x) =   
\frac{ C_{\frac{3}{2}}\, (x^{-1}-A)^{-2} }
     { {x^2}\big[ \,1+\tfrac{1}{2}(x^{-1}-A)^{-2}\, \big]^{2} }\,.
\label{fxA}
\eeq
We will return to this example in our conclusion.\\


\noindent
{\bf Example 4.\,\,}
We consider the special case $q=2$ of equation (\ref{ftildex}).
The limit $q\to 2$ is singular. To take this limit we replace $A$ by a
parameter $a$ defined as
\beq
A = -\frac{q-2}{q-1}\,\log a.
\label{defa}
\eeq
The conditions $A\geq 0$ and $1<q\leq2$ that we
imposed above now require that $a\geq 1$.
Substituting (\ref{defa}) for $A$ in (\ref{ftildex}) 
and taking the limit $q\to 2^-$ we get, with an obvious change of notation,
\beq
\tilde{f}_a(x) = \frac{C_2\,a}{1+a^2x^2}\,,
\label{ftildex2}
\eeq
which is equal to $aG_2(ax)$. Hence we have found here that the $2$-FT
of $aG_2(ax)$ is independent of $a$, for $a\geq 1$.
This independence is in fact valid for all $a>0$ and
is of course easily demonstrated by direct calculation.


\section{Conclusion}
\label{secconclusion}

The authors of reference \cite{Umarovetal08} and later work on the $q$-FT
have certainly been aware of the
necessity for that transformation to have an inverse.
Reference \cite{UmarovTsallis08} 
deals exclusively with this issue.
However, 
the authors limit 
their investigation to the question
of whether {\it a $q$-Gaussian obtained under the $q$-FT
has a unique preimage in the subspace of $q$-Gaussians}, 
that is, to asking if
$q^\prime(q)$ has an inverse. 
The answer to the question thus narrowed down is affirmative
but has little bearing on the 
problem of the invertibility of the $q$-FT
on a full space of functions.
\vspace{3mm}

Our example 3 appears%
\footnote{With the permission of the present author.} 
in the overview by Tsallis \cite{Tsallis09b},
who pays ample attention there to the problem of invertibility. 
Tsallis proposes to select 
a specific value of the parameter $A$, hence 
a specific member of the family $\tilde{f}_A(x)$,
by prescribing the second moment of that function.
Our view, 
quite apart from other questions that such a procedure may raise, is that
this is an {\it ad hoc\,} fix covering the one-parameter case. 
It does not lift the degeneracy in cases 
-- whose existence we have made plausible --
where a single $q$-Fourier transform is associated with
a many-parameter family or
with a family that depends on an arbitrary function, that is, on a
continuum of parameters.
The presentation of our derivation in section \ref{secnoninvert}
had, precisely, the purpose of showing the extent of the invertibility problem.
\vspace{3mm}

In summary, we have shown that, when considered
on a reasonably large space of functions,
the $q$-modified Fourier transformation
employed in the work by Umarov {\it et al.} \cite{Umarovetal08} 
does not possess an inverse.
As a consequence, that work remains unsuccessful in its attempt of
showing that $q$-Gaussians are attractors under addition and rescaling.
\vspace{3mm}

The numerical search for $q$-Gaussians has dealt with 
statistical models that are less
amenable to analytic treatment than those of references
\cite{Tsallis07,Thistletonetal09,Moyanoetal06,HilhorstSchehr07} 
mentioned above.
This search has in particular focused on the logistic map
and the Hamiltonian Mean Field Model (see \cite{Hilhorst09} for references), 
where various types of scaled 
partial sums have been proposed as candidates
for $q$-Gaussian distributed variables.
In the future still other models will no doubt be examined
with the same purpose.
We leave numerical work of that nature out of the present discussion
because, first, 
strictly mathematically speaking we have nothing to
say about it; and, secondly, 
the interpretation of the numerical results has in each case been 
controversial.
The only relevant corollary of the present note is 
that fitting numerical data with $q$-Gaussians 
cannot be justified on the basis of a $q$-central limit theorem.

\section*{Acknowledgment}

The author thanks Professor Constantino Tsallis for discussions and
correspondence.


\appendix


\begin{thebibliography}{10}


\bibitem{Tsallis88}
C. Tsallis, {\it J.~Stat.~Phys.} {\bf 52} (1988) 479.

\bibitem{GellMannTsallis04}
M.~Gell-Mann and C.~Tsallis eds.,
{\it Nonextensive Entropy -- Interdisciplinary Applications,}
Oxford University Press, Oxford (2004).

\bibitem{Tsallis09}
C.~Tsallis,
{\it Introduction to Nonextensive Statistical Mechanics: Approaching a Complex
  World,} Springer, Berlin (2009), {\it to appear}.

\bibitem{Tsallis07}
C. Tsallis, {\it Workshop on the Dynamics of Complex Systems,} 
Natal, Brazil, 11-16 March 2007. 

\bibitem{Thistletonetal09}
W.J.~Thistleton, J.A.~Marsh, K.P.~Nelson, and C.~Tsallis, 
{\it Central Eur. J. Phys.} {\bf 7} (2009) 387.

\bibitem{Moyanoetal06}
L.G.~Moyano, C.~Tsallis, and M.~Gell-Mann,
{\it Europhys.~Lett.} {\bf 73} (2006) 813.

\bibitem{HilhorstSchehr07}
H.J.~Hilhorst and G.~Schehr,
{\it J. Stat. Mech.} (2007) P06003.

\bibitem{Hilhorst09}
H.J.~Hilhorst,
{\it Brazilian J. Physics\,} {\bf 39 } (2009) 371.

\bibitem{Rodriguezetal08}
C.~Rodriguez, V.~Schw\"ammle, and C.~Tsallis,
{\it J.~Stat.~Mech.} (2008) P09006. 

\bibitem{Umarovetal08}
S.~Umarov, C.~Tsallis, and S.~Steinberg,
{\it Milan J. Math.} {\bf 76} (2008) 307.

\bibitem{Tsallis05}
C.~Tsallis,
{\it Milan J. Math.} {\bf 73} (2005) 145. 

\bibitem{UmarovTsallis07}
S. Umarov and C. Tsallis, 
{\it Complexity, Metastability and Nonextensivity, AIP Conference
  Proceedings\,} Vol. {\bf 965} p. 34, 
eds. S.~Abe, H.J.~Herrmann, P.~Quarati, A.~Rapisarda, and C.~Tsallis.
American Institute of Physics, New York (2007).  


\bibitem{Umarovetal10}
S.~Umarov, C.~Tsallis, M. Gell-Mann, and S.~Steinberg,
{\it J.~Math.~Phys.} {\bf 51} (2010) 033502.

\bibitem{UmarovTsallis08}
S.~Umarov and C.~Tsallis,
{\it Phys.~Lett.~A\,} {\bf 372} (2008)4874.

\bibitem{UmarovDuarteQueiros10}
S.~Umarov and S.M.~Duarte~Queir\'os,
{\it J.~Phys.~A\,}  {\bf 43} (2010) 095202.

\bibitem{Tsallisetal09}
C.~Tsallis, A.R.~Plastino, and R.F.~Alvarez-Estrada,
{\it J. Marh. Phys.} {\bf 50} (2009) 043303.

\bibitem{JaureguiTsallis10}
M.~Jauregui and C.~Tsallis,
{\it J. Math. Phys.} {\bf 51} (2010) 063304. 

\bibitem{Tsallis09b}
C. Tsallis, 
{\it Brazilian J. Physics\,} {\bf 39} (2009) 337.

\bibitem{TsallisTirnakli10}
C.~Tsallis and U.~Tirnakli,
{\it Journal of Physics: Conference Series} {\bf 201} (2010) 012001.


\end{thebibliography}
\end{document}